\documentclass[journal]{IEEEtran}
\usepackage{amsmath,amsfonts}
\usepackage{algorithmic}
\usepackage{algorithm}
\usepackage{array}
\usepackage[caption=false,font=normalsize,labelfont=sf,textfont=sf]{subfig}
\usepackage{textcomp}
\usepackage{stfloats}
\usepackage{url}
\usepackage{verbatim}
\usepackage{graphicx}
\usepackage{cite}
\usepackage{amssymb}
\makeatletter

\newcommand{\Rmnum}[1]{\expandafter\@slowromancap\romannumeral #1@}
\makeatother
 
\usepackage{hyperref}
\usepackage{xcolor}
\hypersetup{
    colorlinks=true,        % 启用彩色超链接
    linkcolor=blue,
    citecolor=blue,         % 文献引用颜色
    filecolor=blue,         % 文件超链接颜色
    urlcolor=blue,          % URL 链接颜色
    pdfborder={0 0 0}         % 边框样式（0 0 边框宽度）
}

\usepackage{makecell}
\usepackage{float}
\usepackage{tabularx}
\usepackage{threeparttable}
\usepackage{placeins}
\usepackage{afterpage}

\begin{document}

\title{Multi-User Generative Semantic Communication with \\Intent-Aware Semantic-Splitting Multiple Access}

\author{Jiayi Lu, Wanting Yang, Zehui Xiong,~\IEEEmembership{Senior Member,~IEEE}, Rahim Tafazolli,~\IEEEmembership{Fellow,~IEEE}, \\Tony Q.S. Quek,~\IEEEmembership{Fellow,~IEEE}, and Mérouane Debbah,~\IEEEmembership{Fellow,~IEEE}, Dong In Kim,~\IEEEmembership{Life Fellow,~IEEE}}
        % <-this % stops a space
%\thanks{This paper was produced by the IEEE Publication Technology Group. They are in Piscataway, NJ.}% <-this % stops a space
%\thanks{Manuscript received April 19, 2021; revised August 16, 2021.}}



\maketitle

\begin{abstract}
With the booming development of generative artificial intelligence (GAI), semantic communication (SemCom) has emerged as a new paradigm for reliable and efficient communication. This paper considers a multi-user downlink SemCom system, using vehicular networks as the representative scenario for multi-user content dissemination. To address diverse yet overlapping user demands, we propose a multi-user Generative SemCom-enhanced intent-aware semantic-splitting multiple access (SS-MGSC) framework. In the framework, we construct an intent-aware shared knowledge base (SKB) that incorporates prior knowledge of semantic information (SI) and user-specific preferences. Then, we designate the common SI as a one-hot semantic map that is broadcast to all users, while the private SI is delivered as personalized text for each user. On the receiver side, a diffusion model enhanced with ControlNet is adopted to generate high-quality personalized images. To capture both semantic relevance and perceptual similarity, we design a novel semantic efficiency score (SES) metric as the optimization objective. Building on this, we formulate a joint optimization problem for multi-user semantic extraction and beamforming, solved using a reinforcement learning-based algorithm due to its robustness in high-dimensional settings. Simulation results demonstrate the effectiveness of the proposed scheme.
\end{abstract}

\begin{IEEEkeywords}
Generative semantic communication, diffusion model, semantic information splitting, semantic efficiency
\end{IEEEkeywords}

\section{Introduction}
\IEEEPARstart{W}{ith} the advent of the upcoming sixth-generation (6G) era, cutting-edge applications such as the metaverse, intelligent transportation systems, and smart cities are rapidly emerging. These applications primarily involve multi-user interactions, resulting in an unprecedented surge in data traffic and posing significant challenges to conventional communication systems. Fortunately, semantic communication (SemCom) has emerged as a promising solution, achieving exceptional transmission efficiency while preserving the accuracy of transmitted semantic information (SI) \cite{yang2022semantic}.

Since the emergence of artificial intelligence (AI) technologies, significant efforts have been devoted to the design of SemCom systems. The existing SemCom frameworks can be broadly classified into two paradigms. Most research employs classic joint source-channel coding (JSCC) with end-to-end training\cite{xie2021deep}. However, JSCC limits explainability and flexibility in managing SI and relies on empirical datasets that require costly data collection and labor-intensive labeling. In multi-user scenarios, such limitations become even more pronounced and pose significant challenges to the deployment of end-to-end JSCC framework\cite{10972177}.

The other paradigm focuses on independently training semantic encoding and decoding, which align with modern digital communication systems while simplifying deployment on legacy hardware \cite{han2025scsc}. In particular, this paradigm has been enhanced by advances in generative artificial intelligence (GAI)\cite{zhao2024generative}. Compared to deep learning-based end-to-end SemCom paradigms, GAI enables more diverse forms of SI and highly flexible extraction methods. These advancements facilitate smoother integration with advanced communication technologies. Simultaneously, the user-centric nature of GAI, guided by prompts to generate personalized content, makes it particularly well-suited for multi-user scenarios. It efficiently caters to diverse individual needs, further enhancing its applicability. 

Moreover, current research in SemCom primarily focuses on two key directions: improving the design of semantic encoders and decoders and optimizing the transmission of SI. The former improves SemCom performance by refining semantic encoding and decoding networks, with advancements such as multi-scale semantic feature extraction using the Swin Transformer \cite{pan2023image}, attention-based systems for speech transmission \cite{weng2021semantic}, and generative adversarial network inversion for enhanced transmission robustness \cite{tang2024evolving}. The latter integrates physical-layer techniques, including beamforming in massive multiple-input multiple-output (MIMO) \cite{wu2024deep} and multiple access strategies for resource allocation \cite{yang2023energy, zhang2024scan}. Nevertheless, existing studies often treat semantic extraction and transmission as separate components, overlooking their high interdependence in achieving overall communication performance. Optimizing each component separately makes it challenging to achieve optimal performance for the entire system. Additionally, existing research has overlooked multi-user communication challenges. For example, in metaverse scenarios, users may have different preferences but share the same road infrastructure, leading to unnecessary duplication of common information.

To address the above issues, we develop a joint intent-aware semantic extraction method tailored for the multi-user scenario. Specifically, we assume that the shared knowledge base (SKB) maintains user preferences and personalized intents across different tasks \cite{yi2023deep}. Furthermore, by leveraging pre-trained language models such as BERT\cite{kenton2019bert} and GPT-4\cite{achiam2023gpt}, intent-aware semantic parsing can be effectively implemented to enable flexible semantic extraction and aggregation. Based on this, we propose a framework for multi-user communication systems: Generative SemCom-enhanced intent-aware semantic-splitting multiple access (SS-MGSC). Considering the potential redundancy in information required by multiple users, this approach uniquely enables the simultaneous extraction of a shared SI component along with multiple personalized auxiliary semantic representations. Incidentally and inherently, this approach aligns with the principles of rate-splitting multiple access (RSMA) \cite{mao2022rate}, allowing us to leverage RSMA to further enhance transmission efficiency. To demonstrate the effectiveness of the proposed SS-MGSC framework, we take Vehicle-to-Everything (V2X) as a representative example. In our earlier work\cite{10934748}, we proposed a generative multi-modal SemCom framework in IoV, enhancing semantic encoding, SI transmission, and semantic decoding. Although these studies have advanced the field, a single vehicle's perception is often limited by occlusions and distant objects in real-world scenarios\cite{liu2024select2col}. Therefore, this paper extends our previous work, and the main contributions are summarized as follows:

\begin{itemize}
\item[$\bullet$] We propose an SS-MGSC framework. Specifically, at the transmitter side, we employ two semantic extraction modules: one is to extract common SI in the form of a one-hot map, which offers high robustness in wireless transmission, and the other is to extract personalized SI in the form of text. Meanwhile, we adopt a diffusion model (DM), enhanced with ControlNet, to achieve personalized high-quality image generation.

\item[$\bullet$] 
Given that SemCom prioritizes semantic accuracy over bit error rate (BER) we define a novel semantic metric, the semantic efficiency score (SES), which simultaneously assesses semantic relevance and quantifies perceptual similarity by capturing feature-level differences. To enhance overall system performance under this new metric, we formulate a joint optimization problem that integrates multi-user SI extraction and beamforming. Departing from conventional RSMA frameworks, we substitute typical signal-to-noise ratio (SNR) or channel capacity constraints with SES-based constraints tailored to each user. This approach directly supports our core objective: avoiding over-provisioning for semantic accuracy and optimizing resource efficiency in semantic transmission.

\item[$\bullet$] 
The formulated problem reveals a fundamental trade-off between the volume of SI and the quality of transmission. Due to the implicit and complex relationship between these two metrics, traditional optimization methods are not applicable. To address this, we adopt a reinforcement learning (RL) approach to solve the problem. We validate the effectiveness of the proposed system through extensive simulations, with particular emphasis on the robustness of the one-hot map representation. Results demonstrate that the proposed SS-MGSC framework maintains high-quality transmission even under low-power conditions.
\end{itemize}

The rest of this paper is organized as follows. In Section \Rmnum{2}, we review the state-of-the-art research in RSMA and existing multi-user SemCom research. In Section \Rmnum{3}, we detail the proposed SS-MGSC architecture. Then, in Section \Rmnum{4}, we introduce our proposed semantic efficiency metric and formulate the corresponding optimization problem. Next, in Section \Rmnum{5}, we employ RL algorithm to solve the problem. Simulation results are provided in Section \Rmnum{6}, and we give the conclusions of the study in Section \Rmnum{7}.

\textbf{\textit{Notations:}} Boldface lower symbols represent vectors. $\left\{x_{k}\right\}$ refers to $\left\{x_{1},x_{2},\dots,x_{K}\right\}_{\forall k \in \mathcal{K}}$, where $\mathcal{K}=\{1,2,\dots,K\}$ represents as the set of users. The Hermitian operators is denoted as $(\cdot)^{\mathrm{H}}$, while $\cup$ denotes the union operator. The notation $\mathbb{C}^{M_1 \times M_2}$ represents the space of $M_1 \times M_2$ complex-valued matrices. Morever, $\mathcal{N}[.,.]$ denotes a Gaussian probability distribution, and $\mathbb{E}[.]$ represents the expectation operator.

\section{System Model}
\begin{figure}[!t]
    \centering
    \includegraphics[scale=0.44]{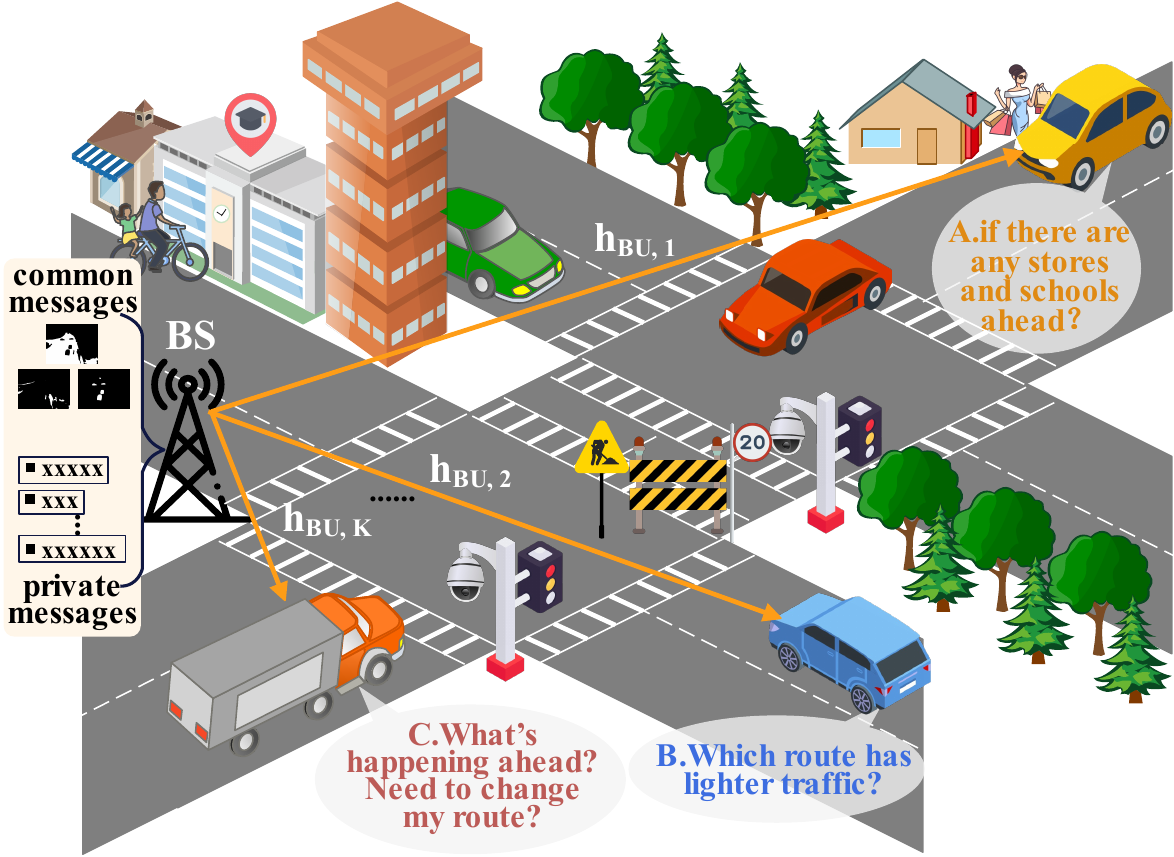}  
    \caption{System model of multi-user Generative SemCom in vehicular network.}
    \label{System model of Generative semantic communication in IoV}
\end{figure}
\begin{figure*}[!t]
    \centering
    \includegraphics[scale=0.52]{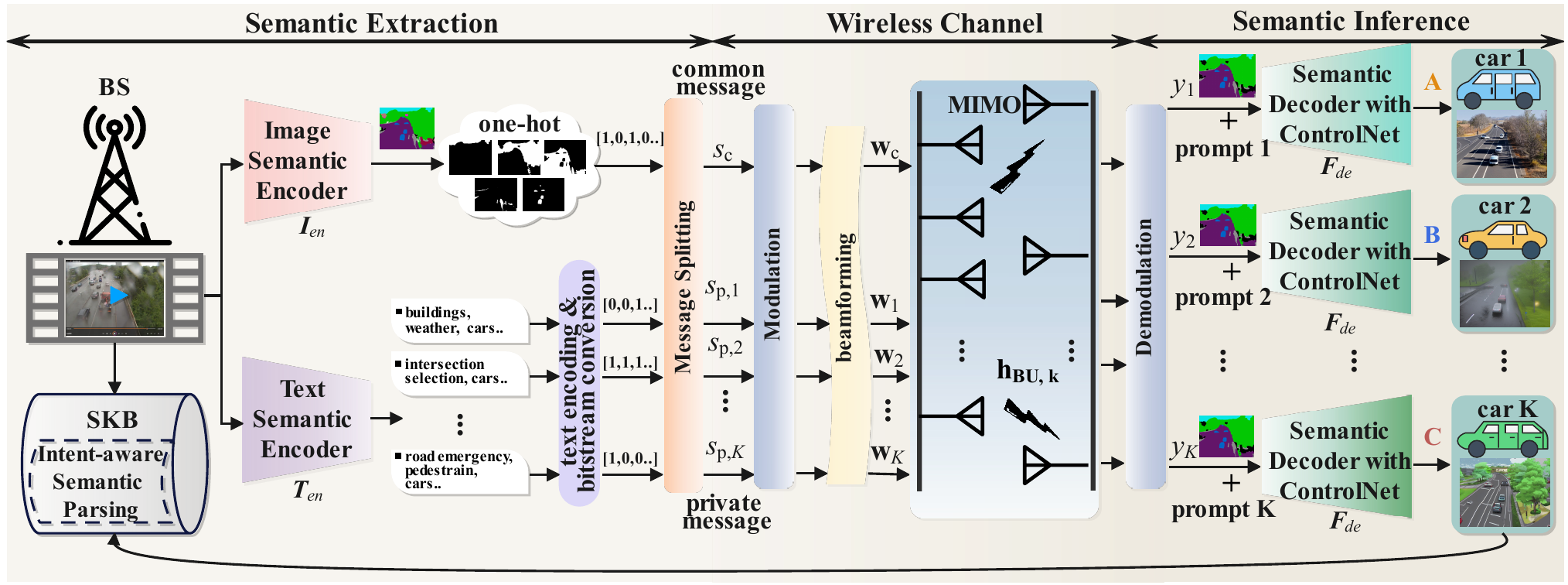}  
    \caption{The detailed architecture of the Intent-aware SS-MGSC.}
    \label{The detailed architecture of the multi-user Generative SemCom with ISTE}
\end{figure*}

\subsection{System overview}
As shown in Fig.~\ref{System model of Generative semantic communication in IoV}, we consider a multi-user downlink SemCom system in vehicular networks. We refer to vehicles as users. Since users have diverse informational needs, different types of contents are prioritized accordingly. In this work, we consider the following three typical categories of users:
\begin{itemize}
\item[\textbf{\textit{A.}}] Remote vehicles: The users with the aim to identify and analyze surrounding community of interest such as nearby buildings, local shops, educational institutions, and current  weather conditions. 
\item[\textbf{\textit{B.}}] Approaching vehicles: The users navigating complex road structures, such as intersections or highway merges, primarily focus on real-time traffic conditions for optimal route selection. In contrast, static environmental details may be redundant for them.
\item[\textbf{\textit{C.}}] Visibility-impaired vehicles: The users in short-range scenarios with restricted visibility, caused by occlusions, require timely updates on sudden events occurring ahead to enhance situational awareness.
\end{itemize}

With the above in mind, to optimize spectral efficiency and minimize redundancy in multi-user content distribution, we propose a strategic approach inspired by RSMA\cite{cheng2023resource}. We assume that the intent-aware SKB has prior knowledge of the SI and individual preferences of each user. The BS broadcasts universally relevant common SI to all users, while simultaneously delivering user-specific private SI to individual users. In our work, the common SI comprises essential road layout information shared among all users, whereas the private SI captures personalized details tailored to individual user needs. At the receiver side, each user utilizes a semantic decoder to reconstruct the transmitted content: first by decoding and subtracting the common SI, then decoding their respective private SI, while treating any irrelevant information as interference to enhance spectrum utilization. The detailed generative SemCom framework is depicted in Fig.~\ref{The detailed architecture of the multi-user Generative SemCom with ISTE}. In the highly dynamic wireless environment of vehicular networks, simplifying the communication process is crucial for enhancing system performance. Thus, in the following, our design is developed in accordance with this objective.

\begin{figure}[!t]
    \centering
    \includegraphics[scale=0.70]{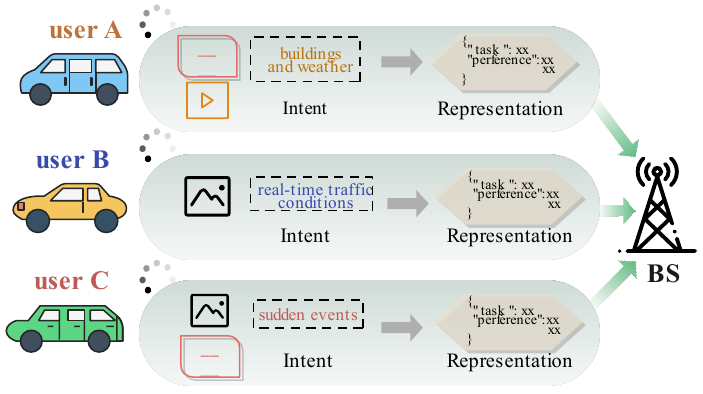}  
    \caption{Construction of Intent-aware SKB.}
    \label{An diagram of Intent-aware SKB}
\end{figure}

To support the proposed framework, the construction of a intent-aware SKB is essential. Traditional deep learning-based SKB predominantly rely on empirical datasets, necessitating extensive data collection and labor-intensive annotation, which not only increases data acquisition costs but also limits system generalization. However, within the framework of GSC, the limitation can be effectively mitigated. As depicted in Fig.~\ref{An diagram of Intent-aware SKB}, we introduce an intent-aware SKB that furnishes global background knowledge, thereby guiding the semantic encoding and decoding process while substantially minimizing reliance on costly data collection and labor-intensive annotation in multi-user scenarios. We first collect user preference information, encompassing multi-modal data such as text, video, and audio across different users.
Then, cutting-edge sequential reasoning large language models\cite{zhao2023survey} are leveraged to integrate this information into a unified semantic representation. Finally, the extracted SI is structured and stored within the knowledge base. 

\subsection{Semantic Information Extraction}
Building upon the intent-aware SKB, we consider the semantic encoder at the transmitter that consists of two semantic extraction modules. For common SI, we adopt the image modality and employ a visual semantic extraction module, denoted by $\boldsymbol{I}_{en}(\cdot)$, which extracts semantic segmentation information to represent the road structure. For private SI, we use the text modality, which contains user-specific and more detailed content.
The text semantic extraction module, denoted by $\boldsymbol{T}_{e n}(\cdot)$, is used to capture individual user preferences. Accordingly, the BS simultaneously transmits image and textual information. Consequently, the SI extracted by the $k$th user can be represented by
\begin{equation}
    \begin{aligned}
        \label{semantic encoder}
        s_k=\boldsymbol{I}_{e n}\left(\mathcal{C}, n_c\right) \cup \boldsymbol{T}_{e n}\left(\mathcal{C}, n_{p, k}\right),
    \end{aligned}
\end{equation}
where $\mathcal{C}$ represents source data, $n_c \in \mathcal{M} = \{0,1, 2, 3, ..., M\} $ and $n_{p,k} \in \mathcal{N} = \{0, 1, 2, 3, ..., N\}$ represent the number of units for common data and private data for the $k$th user, respectively.

The image semantic encoder can utilize UPerNet \cite{xiao2018unified}, or SegFormer \cite{xie2021segformer} to generate semantic segmentation maps, which are then classified into different semantic classes based on predefined labels. To enhance robustness in noisy transmission environments, as demonstrated in the simulation results presented in Section \Rmnum{5}, the segmentation maps are subsequently converted into one-hot encoded representations. Additionally, one-hot encoded maps can further reduce the volume of transmitted data, thereby significantly enhancing transmission efficiency. Meanwhile, the text semantic encoder utilizes prompt inversion \cite{mahajan2024prompting} as a soft prompt for the generative model, while selectively incorporating hard prompts to further refine its performance \cite{wen2023hard}. Then, the refined text prompt is encoded into a bitstream using ASCII encoding for digital transmission.

\subsection{Semantic Information Transmission}
We consider a multi-user MIMO downlink SemCom system in vehicular networks, where a BS equipped with $N_t$ antennas serves $K$ single-antenna vehicles.
After SI extraction and splitting, the data stream is collectively denoted as $\mathbf{s}=\left[s_1, s_2, \ldots, s_K\right]^{\top}$. In highly dynamic wireless channels such as vehicular communications, the number of symbols per SI is a key design parameter to maintain robustness subject to the total transmit power at BS. Namely, adaptive selection of M is desired when the channel gain fluctuates wildly over time due to the fast mobility of vehicles. 

To ensure efficient multi-user communication, we employ the MIMO technology, where both modulated data streams are linearly precoded using the precoding matrix $\mathbf{W}=[\overline{\mathrm{w}}_1, \overline{\mathrm{w}}_2,\ldots, \overline{\mathrm{w}}_K]$. The transmitted signal from the transmitter can be expressed as
\begin{equation}
    \begin{aligned}
        \label{beamforming}
        \mathbf{x} = \mathbf{W}\mathbf{s}.
    \end{aligned}
\end{equation}

For the $k$th user, the received signal can be represented as
\begin{equation}
    \begin{aligned}
        \label{sssignal}
        y_k=\mathbf{h}_{\mathrm{BU}, k}^\mathrm{H} \mathbf{x}+z_k = \mathbf{h}_{\mathrm{BU}, k}^\mathrm{H}\overline{\mathrm{w}}_k\mathbf{s}_k +\sum_{\substack{l \ne k}}^K  \mathbf{h}_{\mathrm{BU}, k}^\mathrm{H} \overline{\mathrm{w}}_l\mathbf{s}_l+z_k,\\
    \end{aligned}
\end{equation}
where $z_k \sim \mathcal{C N}\left(0, \sigma^2\right)$ denotes the additive white Gaussian noise (AWGN) received by $k$th user and the second term in (\ref{sssignal}) represents the multi-user interference (MUI). Given that vehicular communication typically experiences a highly dynamic wireless environment with rich scattering, we apply a Rayleigh fading channel to model the small-scale fading environment. The channel between the BS and the $k$th user is represented by $\mathbf{h}_{\mathrm{BU}, k}=\sqrt{\epsilon_0\left(\frac{d}{d_0}\right)^{-\alpha}} \mathbf{\tilde{h}}_{\mathrm{BU}, k}$, where $\mathbf{h}_{\mathrm{BU}, k}\in \mathbb{C}^{N_t \times 1}$, $d$ is the distance, $\epsilon_0\left(\frac{d}{d_0}\right)^{-\alpha}$ is the large-scale path loss for the reference distance $d_0$, and $\mathbf{\tilde{h}}_{\mathrm{BU}, k}$ represents the small-scale fading, which follows a Rayleigh fading model.

\subsection{Semantic Decoding}
Research on one-step diffusion has become increasingly mature in recent studies \cite{wang2024rectified, liu2023instaflow}, significantly reducing computational requirements without compromising the quality of generated images. With the support of this technology, each vehicle has sufficient computational capacity\cite{lu2024generative} and is equipped with a generative model serving as the semantic decoder $\boldsymbol{F}_{de}(\cdot)$ at the receiver. To further enhance the fidelity of the reconstructed images, we assume that the receiver employs a pre-trained ControlNet model integrated with Stable DM, leveraging its ability to refine structural details and improve visual quality compared to using Stable DM alone. At this point, the mathematical relationship between the signal received by the $k$th user and the generated output image can be expressed as
\begin{equation}
    \begin{aligned}
        \label{semantic decoder}
        \hat{\mathcal{C}}_{k} = \boldsymbol{F}_{de}(y_k).
    \end{aligned}
\end{equation}

\section{Problem Formulation}
Based on the proposed SS-MGSC framework, we introduce a semantic effectiveness metric in Section \Rmnum{4-A} to evaluate SemCom performance. Then, taking this metric as the optimization objective, we formulate an optimization problem in Section \Rmnum{4-B} that jointly optimize transmission beamforming and the semantic extraction.

\subsection{Semantic Efficiency Score}
\label{sec:SES}
Unlike traditional communication, which evaluates performance based on bit-level accuracy, SemCom requires new metrics to assess transmission effectiveness at the semantic level. Moreover, given the concurrent transmission of image and text in our scenario, evaluating the semantic similarity separately, as commonly adopted in prior SemCom communication, fails to capture holistic semantic effectiveness. Thus, we define a novel semantic evaluation metric, SES, to measure the effectiveness of SemCom. SES is constructed of contrastive language-image pre-training (CLIP) \cite{radford2021learning} and learned perceptual image patch similarity (LPIPS) \cite{zhang2018unreasonable}, providing a quantitative measure of semantic fidelity. More precisely, CLIP leverages large-scale vision-language pretraining to capture the semantic relevance between generated images and the text descriptions, while LPIPS quantifies the perceptual similarity by measuring feature-level differences between images. Given the involvement of both text and image modalities in our scenario, where both semantic content and perceptual quality are critical, the proposed SES incorporates semantic similarity and perceptual structure to better evaluate the effectiveness of SemCom. Additionally, compared to metrics like structural similarity index measure (SSIM) and peak signal-to-noise ratio (PSNR) \cite{hore2010image}, which focus on pixel-level similarity, our proposed SES metric better capture the semantic content of images, better aligning with the considered communication task. A higher SES value indicates that richer and more relevant SI has been successfully received, potentially enhancing the visual quality of the generated images.

We calculate CLIP by the cosine similarity between the prompt truth $\mathcal{C}_{T}$ and the synthesized images $\{\hat{\mathcal{C}}_{k}\}_{\forall k \in \mathcal{K}}$, capturing the semantic alignment between the text and the generated image. Since the original CLIP score is defined within the range of $[-1,1]$, we transform it to $[0,1]$ to facilitate subsequent computations. Additionally, we utilize LPIPS between orignal image $\{\mathcal{C}_{k}\}_{\forall k \in \mathcal{K}}$ and the synthesized images $\{\hat{\mathcal{C}}_{k}\}_{\forall k \in \mathcal{K}}$ to assess the structural and perceptual fidelity of the images, leveraging a pre-trained VGG network to quantify human perceptual similarity. These metrics can be written as
\begin{equation}
    \label{CLIPS}
    \operatorname{CLIP}(\mathcal{C}_{T}, \hat{\mathcal{C}}_{k}) = 
    \frac{\mathbb{E} \left[ 
    \frac{ f(\mathcal{C}_{T}) \cdot f(\hat{\mathcal{C}}_{k}) } 
         { \left\| f(\mathcal{C}_{T}) \right\| \cdot \left\| f(\hat{\mathcal{C}}_{k}) \right\| } 
    \right] + 1}{2},
\end{equation}
\begin{equation}
    \begin{aligned}
        \label{LPIPS}
        \operatorname{LPIPS}(\mathcal{C}_{k}, \hat{\mathcal{C}}_{k}) = \sum_l \frac{1}{{H_l} {W_l}} \sum_{h, w}^{{H_l}, {W_l}}\left\|\eta_l \odot\left(y_{h, w}^l-\hat{y}_{h, w}^l\right)\right\|_2^2,
    \end{aligned}
\end{equation}
where the height and width of the output feature map from the $l$-th VGG layer are denoted as $H_l$ and $W_l$, respectively. The variables $y_{h, w}^l$ and $\hat{y}_{h, w}^l$ represent the $\left(h,w\right)$-th element of the output feature maps for $\mathcal{C}_{k}$ and $\hat{\mathcal{C}}_{k}$. Furthermore, the weight $\eta_l$ regulates the contribution of each layer, prioritizing features most aligned with human perceptual relevance. 

Accordingly, we define the SES for each user as $\operatorname{CLIP}_{k} + (1 - \operatorname{LPIPS}_{k})$. By combining CLIP and LPIPS, this matrix ensures a balance between semantic consistency and visual quality, addressing both high-level content alignment and low-level image details. The SES is primarily influenced by the quantity of extracted SI, including both common and private SI, and the power allocation of semantic transmission. 

\subsection{Problem Formulation}
\label{Problem Formulation}
For semantic splitting, the previously introduced data stream $\mathbf{s}$ intended for the $K$ users is further segmented and encapsulated into one common semantic information (SI) component $s_\mathrm{c}$ and several private components ${s_{\mathrm{p}, 1}, s_{\mathrm{p}, 2}, \dots, s_{\mathrm{p}, K}}$. The proposed semantic-splitting multiple access (SSMA) is adopted to reduce redundant information transmission in multi-user communication. Therefore, both modulated data streams are linearly precoded using the precoders $\mathbf{w} = \{\mathbf{w}_\mathrm{c}, \mathbf{w}_1,\dots,\mathbf{w}_K\}$, where $\mathbf{w}_\mathrm{c} \in \mathbb{C}^{N_t \times 1}$ is the common data stream precoder, and $\mathbf{w}_k\in \mathbb{C}^{N_t \times 1}$ denotes the private data stream precoder for vehicle $k$. Consequently, the transmitted signal from the BS can be expressed as the superposition of the common SI stream and the sum of the private SI streams:
\begin{equation}
    \begin{aligned}
        \label{beamforming}
        \mathbf x = \sum_{k=1}^K \mathbf{w}_k s_{\mathrm{p}, k}+\mathbf{w}_\mathrm{c} s_\mathrm{c}.
    \end{aligned}
\end{equation}

We use the letter $z$ to represent the AWGN. Accordingly, the $k$th user's received signal can be expressed:
\begin{equation}
    \begin{aligned}
        \label{received signal}
        y_k=\mathbf{h}_{\mathrm{BU}, k}^\mathrm{H} \mathbf{x}+z_k = \sum_{j=1}^K \mathbf{h}_{\mathrm{BU}, k}^\mathrm{H} \mathbf{w}_j s_{\mathrm{p}, j}+\mathbf{h}_{\mathrm{BU}, k}^\mathrm{H} \mathbf{w}_\mathrm{c} s_\mathrm{c}+z_k,\\
    \end{aligned}
\end{equation}
where the term$\sum_{j=1, j \neq k}^K \mathbf{h}_{\mathrm{BU}, k}^\mathrm{H} \mathbf{w}_j s_{\mathrm{p}, j}$ corresponds to the MUI in (\ref{received signal}) for the $k$-th user.

The transmission quality of different SI is primarily determined by the SINR. For the $k$th user, the SINR for transmitting the common SI, denoted as $\gamma_{\mathrm{c}, k}$, and the private SI, denoted as $\gamma_{\mathrm{p}, k}$, can be expressed separately as follows:
\begin{equation}
    \begin{aligned}
        \label{rate1}
        &\gamma_{\mathrm{c}, k}=\frac{\left|\mathbf{h}_{\mathrm{BU}, k}^\mathrm{H}\mathbf{w}_\mathrm{c}\right|^2}{\sum_{j=1}^K \left|\mathbf{h}_{\mathrm{BU}, k}^\mathrm{H}\mathbf{w}_j\right|^2+\sigma^2},
    \end{aligned}
\end{equation}
\begin{equation}
    \begin{aligned}
        \label{rate2}
        \gamma_{\mathrm{p}, k}=\frac{\left|\mathbf{h}_{\mathrm{BU}, k}^\mathrm{H}\mathbf{w}_k\right|^2}{\sum_{j=1, j \neq k}^K\left|\mathbf{h}_{\mathrm{BU}, k}^\mathrm{H}\mathbf{w}_j\right|^2+\sigma^2}.
    \end{aligned}
\end{equation}

Based on the preceding definitions, our objective is to maximize the SES to enhance the quality of generated images while adhering to several constraints. Since we consider the transmission of semantic representations and account for transmission distortions, traditional Shannon-based capacity formulations are no longer applicable. Accordingly, both transmission quality and semantic extraction are jointly incorporated into the objective function, which can be expressed as $f_{k}\left(n_\mathrm{c},n_{\mathrm{p},k}, \mathrm{SINR}_c, \mathrm{SINR}_{\mathrm{p},k} \right) = \operatorname{CLIP}_{k} + (1 - \operatorname{LPIPS}_{k})$.

Mathematically, we formulated a maximization problem as
\begin{subequations}
    \begin{align}
        %\label{CLIP}
        \displaystyle \max_{\mathbf{w}, {n_\mathrm{c}}, {n_{\mathrm{p},k}}} 
        & \sum_{k=1}^K f_{k}\left(n_\mathrm{c},n_{\mathrm{p},k}, \mathrm{SINR}_c, \mathrm{SINR}_{\mathrm{p},k} \right), \label{obj_ori}\\
        \text{s.t.} & \sum_{k=1}^K\left\|\mathbf{w}_k\right\|_{\mathrm{2}}^2 + \left\|\mathbf{w}_{\mathrm{c}}\right\|_{\mathrm{2}}^2  \leq P_{\max}, \label{st_ori_P}\\
        & f_{k}\left(n_\mathrm{c},n_{\mathrm{p},k}, \mathrm{SINR}_c, \mathrm{SINR}_{\mathrm{p},k} \right) \geq I_{\mathrm{th},k},\quad \forall k \in \mathcal{K},\label{st}\\
        & n_\mathrm{c} = \{0,1, 2, 3, ..., M\},\label{snc}\\
        & n_{\mathrm{p},k} =  \{0,1, 2, 3, ..., N\},\quad \forall k \in \mathcal{K},\label{snt}
    \end{align}
\end{subequations}
where (\ref{st_ori_P}) imposes a limit under maximum transmit power $P_{\max}$, constraint (\ref{st}) ensures that each user's SES meets the threshold $I_{\mathrm{th},k}$ to guarantee the quality of the generated image. Constraints (\ref{snc}) and (\ref{snt}) specify the number of common SI one-hot images and private SI prompts that can be transmitted based on available resources. 

Since SemCom focuses on semantic accuracy and the volume of SI, rather than channel capacity and BER as reflected in the objective (\ref{obj_ori}), we have removed the Shannon formula-based channel capacity constraints from conventional RSMA. Instead, we directly constrain the SES. The optimization aims to maximize SES, which is influenced by both the quantity of extracted SI and the quality of SI transmission. On the one hand, increasing the transmission of common and private SI enhances the total volume of SI. However, due to power constraints (\ref{st}), the per-symbol power allocation is reduced potentially lowering SINR and degrading transmission quality. On the other hand, reducing the volume of SI transmission increases per-symbol power allocation but may result in insufficient SI, ultimately impacting SES. Beyond the aforementioned trade-off, the lack of a closed-form expression for the objective function exacerbates the complexity of this problem.

\section{Deep Reinforcement Learning-based Framework Design}
From the formulated optimization problem, it is evident that the problem is NP-hard, characterized by both continuous and discrete variables, making it difficult to obtain the optimal solution directly. To address this challenge, we reformulate the original problem as a Markov Decision Process (MDP) and employ the RL approach, leveraging the agent to learn optimal policies through interactions with the environment, thereby effectively tackling the problem.

\subsection{MDP Formulation}
We first model the wireless communication network as the environment, with the agent located at the BS, which is responsible for jointly determining multi-user semantic extraction and power allocation. The formulated MDP consists of an agent and a four-tuple $\langle\mathcal{S}, \mathcal{A}, r, \gamma \rangle$, which characterizes the way the BS interfaces with the broader network system. Specifically, $\mathcal{S}$ represents the state space, $\mathcal{A}$ denotes the action space, $r$ denotes the instantaneous reward at the end of each step, and $\gamma$ denotes the discount coefficient that influences the weight assigned to future rewards during the learning process. The agent progressively learns to establish a mapping between states and actions, aiming to maximize the cumulative reward over time. In the following, we provide a detailed description of the four-tuple components established for the formulated optimization problem.

\textit{\text {1)} \textbf{\textit{State Space:}}}
The state space is designed to encapsulate as much relevant environmental information as possible to facilitate solving the problem. Specifically, we consider the channel conditions and certain threshold values as key environmental factors. Morever, since neural networks, such as those implemented in PyTorch, do not inherently support complex-valued computations, We represent complex values by separating them into imaginary and real parts.

\begin{equation}
    \begin{aligned}
        \label{state space}
        \mathcal{S} = \{ \operatorname{Re}\{\mathbf{h}_{\mathrm{BU}, k}\},\operatorname{Im}\{\mathbf{h}_{\mathrm{BU}, k}\}, P_{\max }, I_{\mathrm{th},k}\}_{\forall k \in \mathcal{K}},
    \end{aligned}
\end{equation}
where the dimensionality of the state space is given by $(2N_tK+1+K)$.

\textit{\text {2)} \textbf{\textit{Action Space:}}}
The agent perceives the current environment state and chose an action accordingly, which determines the transition to the next state. The action space comprises the common SI beamforming vector $\mathbf{w}_\mathrm{c}$, the private SI beamforming vector $\mathbf{w}_k$, the number of units allocated for common data $n_c$, and the number of units for private data $n_{p,k}$. Similarly, the complex-valued variables $\mathbf{w}_\mathrm{c}$ and $\mathbf{w}_k$ need to be converted into their real and imaginary components before serving as input to neural network. The action space is defined as
\begin{equation}
    \begin{aligned}
        \label{action space}
        \mathcal{A} = \{\mathbf{w}_\mathrm{c}, \{\mathbf{w}_k\},{n_\mathrm{c}}, {n_{\mathrm{p},k}} \}_{\forall k \in \mathcal{K}},
    \end{aligned}
\end{equation}
with the action space is expressed by $(2N_t+2N_tK+1+K)$. 

\textit{\text {3)} \textbf{\textit{Discount Factor:}}} The discount factor $\gamma \in(0,1]$ represents the degree of importance assigned to future rewards. The value of$\gamma$ close to 0 encourages myopic behavior, prioritizing immediate rewards, whereas the value of $\gamma$ close to 1 promotes far-sighted behavior, emphasizing long-term gains while accounting for future uncertainties.

\textit{\text {4)} \textbf{\textit{Reward Function:}}}
The reward function consists of two components: the objective function and a penalty term to enforce constraint satisfaction. To balance the instantaneous reward and the penalty term, we incorporate the weighting factors $\alpha$ and $\beta$ into the reward function, which is defined as follows:
\begin{equation}
    \begin{aligned}
    \label{reward}
    r=&\sum_{k=1}^K f_{k}\left(n_\mathrm{c},n_{\mathrm{p},k}, \mathrm{SINR}_c, \mathrm{SINR}_{\mathrm{p},k} \right) \\
    & + \alpha \min(0, P_{\max} - \sum_{k=1}^K\left\|\mathbf{w}_k\right\|_{\mathrm{2}}^2 + \left\|\mathbf{w}_{\mathrm{c}}\right\|_{\mathrm{2}}^2)\\
    &+ \beta \sum_{k=1}^K\left[ \min(0,  f_{k}\left(n_\mathrm{c},n_{\mathrm{p},k}, \mathrm{SINR}_c, \mathrm{SINR}_{\mathrm{p},k} \right) - I_{\mathrm{th},k})\right].
    \end{aligned}
\end{equation}

In this formulation, the instantaneous reward at each time step $t$ is determined by the system's current state and selected action. This reward provides immediate feedback to the agent, guiding its short-term decisions. Nevertheless, the ultimate aim of reinforcement learning is to maximize the long-term accumulated reward, which is defined as:
\begin{equation}
    \begin{aligned}
        \label{accumulated reward}
        \mathcal{R} = \sum_{l=0}^{\infty} \gamma^k r_{t+l}, 
    \end{aligned}
\end{equation}
where $l$ represents the time step offset from the current time step $t$.

\textit{\text {5)} \textbf{\textit{Policy:}}} The policy $\pi_{\boldsymbol{\theta}}\left(\boldsymbol{s}_t, \boldsymbol{a}_t\right)$ specifies the likelihood of choosing action $a_t$ given the state $\boldsymbol{s}t$. It adheres to the normalization constraint, ensuring that in discrete action spaces, $\sum{\boldsymbol{a}t \in \mathcal{A}} \pi{\boldsymbol{\theta}}\left(\boldsymbol{s}t, \boldsymbol{a}t\right)=1$, or in continuous action spaces, $\int{\mathcal{A}} \pi{\boldsymbol{\theta}}\left(\boldsymbol{s}_t, \boldsymbol{a}_t\right) d \boldsymbol{a}_t=1$, where $\boldsymbol{s}_t \in \mathcal{S}$ and $\boldsymbol{a}_t \in \mathcal{A}$ at each time step $t$. The parameter $\theta$ represents the learnable aspects of the policy, which are optimized through methods like policy gradients. The policy defines the agent's behavior by specifying a probability distribution over possible actions for any given state, steering the agent's interaction with its environment.

\subsection{Proximal Policy Optimization-Based Algorithm}
\begin{figure}[!t]
    \centering
    \includegraphics[scale=0.40]{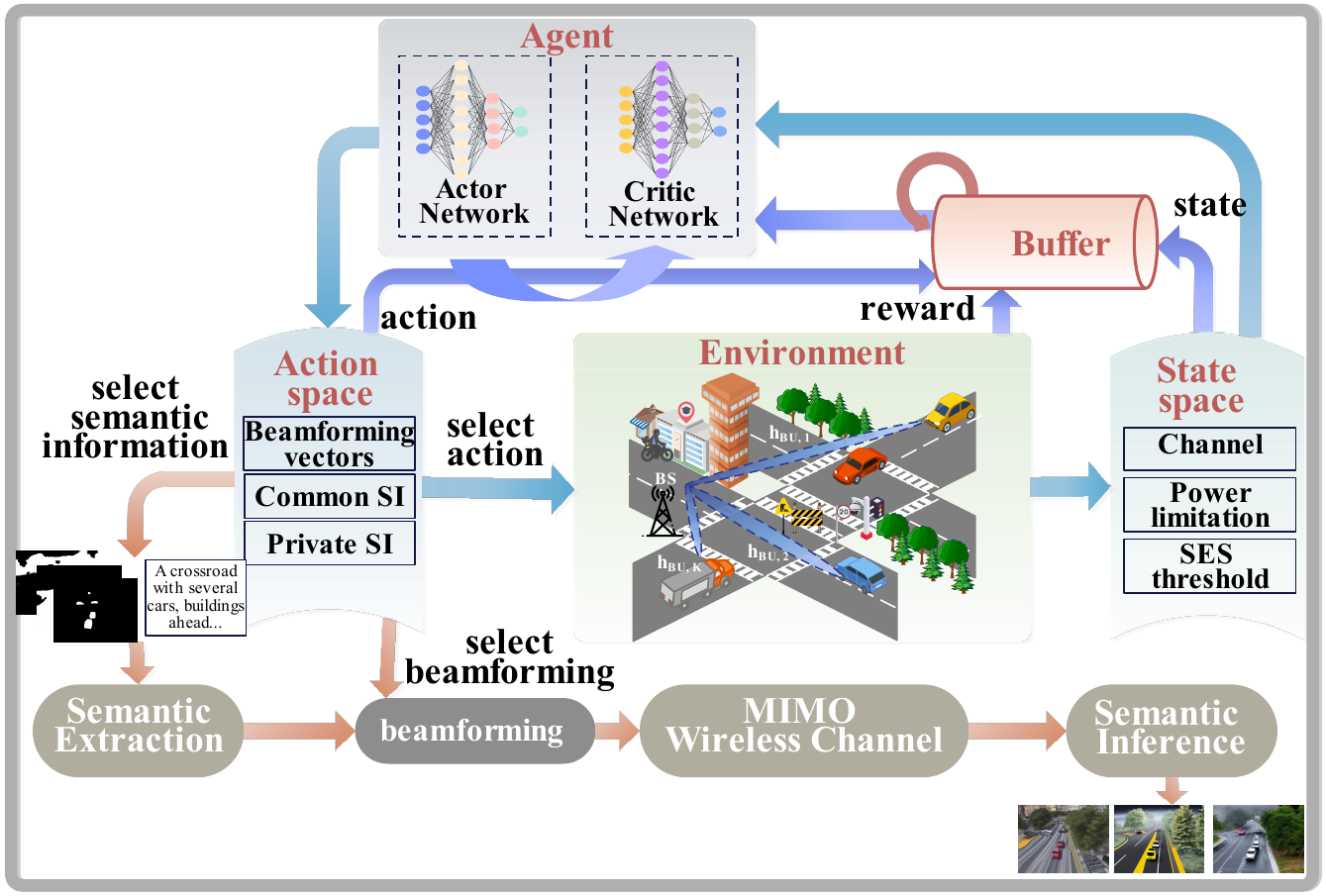}  
    \caption{The PPO-based SS-MGSC framework.}
    \label{PPO}
\end{figure}
Considering its stability and sample efficiency, Proximal Policy Optimization (PPO) \cite{schulman2017proximal}, a policy gradient-based deep RL algorithm, is adopted in this work. In our problem, decomposing complex-valued variables into real and imaginary components increases the dimensionality of the action and state spaces, challenging policy optimization. Given PPO’s robustness in high-dimensional settings, we adopt it to learn optimal policies efficiently, and detailed processes are demonstrated in Fig.~\ref{PPO}.

In the actor-critic framework of PPO, the actor network develops a policy for action selection, and the critic network evaluates the value function to assist in guiding the learning. At timestep $t$, the actor network $\pi_{\boldsymbol{\theta}}$, parameterized by $\theta$, takes the state $\boldsymbol{s}_t$ as input. For discrete variables in our action space, we first apply a continuous relaxation to facilitate gradient-based optimization. After this transformation, the action space becomes a fully continuous action space, thus the actor network models the policy $\pi_{\boldsymbol{\theta}}\left(\boldsymbol{s}_t, \boldsymbol{a}_t\right)$ as a Gaussian distribution, outputting the mean $\boldsymbol{\mu}\left(\boldsymbol{s}_t\right)$ and standard deviation $\boldsymbol{\sigma}\left(\boldsymbol{s}_t\right)$. The agent then samples an $\boldsymbol{a}_t$ from this distribution. Meanwhile, the critic network $V_{\varphi}\left(\boldsymbol{s}_t\right)$, parameterized by $\varphi$, estimates the value of the state, indicating the anticipated cumulative reward from state $\boldsymbol{s}_t$. This estimation is used to evaluate and improve the policy, ensuring more stable and effective learning.
 
Compared to standard policy gradient methods, PPO improves training stability by introducing a clipping mechanism that constrains policy updates, preventing excessive changes that could destabilize learning. The actor network is optimized using the clipped surrogate objective:
\begin{equation}
    \begin{aligned}
    \label{actor CLIP}
    \mathcal{L}^{\text {clip}}(\boldsymbol{\theta})=\hat{\mathbb{E}}_t\left[\min \left(r_t(\boldsymbol{\theta}) \hat{A}_t, \operatorname{clip}\left(r_t(\boldsymbol{\theta}), 1-\epsilon, 1+\epsilon\right) \hat{A}_t\right)\right],
    \end{aligned}
\end{equation}
where $\epsilon$ is a clipping hyperparameter that modifies the surrogate objective function by limiting the impact of large policy updates. Specifically, it discourages updates where the probability ratio $r_t(\boldsymbol{\theta})$ deviates significantly from 1, constraining it within the range $[1-\epsilon, 1+\epsilon]$ and thereby reducing the risk of unstable training. The estimated advantage function $\hat{A}_t$, quantifies the relative benefit of choosing a specific action in comparison to the average action selected by the policy in state $\boldsymbol{s}_t$ at time $t$.

Specifically, $r_t(\boldsymbol{\theta})=\frac{\pi_{\theta_{\text {new}}}\left(\boldsymbol{a}_t \mid \boldsymbol{s}_t\right)}{\pi_{\theta_{\text {old }}}\left(\boldsymbol{a}_t \mid \boldsymbol{s}_t\right)}$, quantifying the relative likelihood of an action under the new policy compared to the old policy. It serves as a key factor in leveraging experiences collected under the old policy to estimate performance under the updated policy. To improve training stability, PPO commonly adopts Generalized Advantage Estimation (GAE), which balances bias and variance in advantage estimation. This leads to more stable policy updates compared to standard temporal difference methods.
\begin{equation}
    \begin{aligned}
    \label{A_t}
    \hat{A}_t=\delta_t+(\gamma \lambda) \delta_{t+1}+\cdots+(\gamma \lambda)^{T-t+1} \delta_{T-1},
    \end{aligned}
\end{equation}
where the temporal difference residual is given by $\delta_t = r_t + \gamma V_{\boldsymbol{\varphi}}(s_{t+1}) - V_{\boldsymbol{\varphi}}(s_t)$, and $\lambda$ represents the smoothing parameter, controlling the balance between bias and variance in the advantage estimation. The critic network approximates the state value function $V_{\boldsymbol{\varphi}}(s_t)$, which estimates the expected total reward starting from state $\boldsymbol{s}t$ while following the current policy $\pi\left(\boldsymbol{s}t, \boldsymbol{a}t\right)$. Formally, it is defined as: $V{\boldsymbol{\varphi}}(s_t) = \mathbb{E}\pi\left[\sum{k=0}^{\infty} \gamma^k r_{t+k} \mid \boldsymbol{s}_t\right]$.

In addition, the critic network is updated by minimizing the mean squared error between the predicted state value $V_{\varphi}\left(s_t\right)$ and the target value $\hat{V}_t$, which approximates the actual value function:
\begin{equation}
    \begin{aligned}
    \label{critic}
    \mathcal{L}^{\text {critic}}(\boldsymbol{\varphi})=\mathbb{E}_t\left[\left(V_{\boldsymbol{\varphi}}(s_t)-\hat{V}_t\right)^2\right],
    \end{aligned}
\end{equation}
where $\hat{V}_t$ can be computed using either the GAE method as: $\hat{V}_t=\hat{A}_t+V_{\varphi}\left(s_t\right)$.
The GAE method provides a balance between bias and variability in advantage estimation, leading to more stable policy updates.

Accordingly, the PPO-based SES-oriented SSMA optimization is outlined in Algorithm \ref{alg1} for better clarity. The agent undergoes iterative training until it reaches the predefined maximum number of steps.
\begin{algorithm}[H]
\caption{PPO-based SES-oriented SSMA optimization}\label{alg1}
\begin{algorithmic} [1]
\STATE \textbf{Input:} \parbox[t]{0.8\linewidth}{State $\{ \operatorname{Re}\{\mathbf{h}_{\mathrm{BU}, k}\},\operatorname{Im}\{\mathbf{h}_{\mathrm{BU}, k}\}, P_{\max }, I_{\mathrm{th},k}\}_{\forall k \in \mathcal{K}}$.}
\STATE \textbf{Output:} Action $\{\mathbf{w}_\mathrm{c}, \{\mathbf{w}_k\},{n_\mathrm{c}}, {n_{\mathrm{p},k}} \}_{\forall k \in \mathcal{K}}$, and maximized SES.
\STATE \textbf{Initialization:} Actor network parameter $\theta$, critic network parameter $\varphi$, optimizers with learning rates $\alpha_a, \alpha_c$, gradient clipping threshold $\beta_a, \beta_c$, buffer capacity $D$ and mini-batch size $H$.
\STATE \textbf{for} episode = 1 to max episode \textbf{do}
\STATE \hspace{0.5cm} Reset environment and the replay buffer $D$;
\STATE \hspace{0.5cm} \textbf{for} step = 1 to max step \textbf{do}
\STATE \hspace{1cm} Retrieve current state $\boldsymbol{s}_t$;
\STATE \hspace{1cm} \parbox[t]{0.8\linewidth}{Input $\boldsymbol{s}_t$ into actor network and compute action $\boldsymbol{a}_t$ using policy $\pi_{\theta}(\boldsymbol{s}_t)$;}
\STATE \hspace{1cm} Execute $\boldsymbol{a}_t$;
\STATE \hspace{1cm} \parbox[t]{0.8\linewidth}{Observe the next state \(\boldsymbol{s}_{t+1}\) and transition to the next state; calculate the reward \(\boldsymbol{r}_t\).}
\STATE \hspace{1cm} Store transition $\left\{s_t, a_t, s_{t+1}, r_t\right\}$ in $D$;
\STATE \hspace{1cm} \parbox[t]{0.8\linewidth}{\textbf{If buffer is full}, sample a mini-batch $H$ from \(\mathcal{D}\);}
\STATE \hspace{1cm} Compute advantage estimate $\hat{A}_t$ using (\ref{A_t});
\STATE \hspace{1cm} \parbox[t]{0.8\linewidth}{Compute policy ratio $r_t(\theta)$ and clipped surrogate loss using (\ref{actor CLIP});}
\STATE \hspace{1cm} \parbox[t]{0.8\linewidth}{Compute value loss using (\ref{critic});}
\STATE \hspace{1cm} Clear buffer;
\STATE \hspace{1cm} \textbf{end for}
\STATE \hspace{0.5cm} Update total reward;
\STATE \hspace{0.5cm} Save model parameters and update visualization;
\STATE \textbf{end for}
\end{algorithmic}
\label{alg1}
\end{algorithm}

\section{Numerical Results}
In this section, we showcase the experimental results to demonstrate the effectiveness of the SS-MGSC framework. The semantic similarity of the regenerated images is evaluated using the proposed SES metric, while leveraging CLIP for semantic consistency assessment and LPIPS to capture fine-grained visual differences, ensuring a comprehensive evaluation.

\subsection{Simulation Setup}
\subsubsection{Dateset and Experimental Settings}
We use the AAU RainSnow dataset\cite{bahnsen2018rain}, which consists of video captured by an RGB camera mounted on the street lamp at an intersection under various weather conditions, including rain, snow, and general adverse weather for traffic surveillance. In our experiment, we use a rainy road scene as the source data. The image transmitted is 480 $\times$ 640 pixels, approximately 0.88~Mb or 416~KB. The average size of a one-hot map is about ${L}_{c}$~=~3.3~Kb. It is evident that using one-hot encoding significantly reduces the volume of transmitted SI, which is crucial for high-data-demand V2X communication.

We utilize the pre-trained model \hypersetup{pdfborder={0 0 0}}\footnote{https://github.com/ubc-vision/Prompting-Hard-Hardly Prompting}\hypersetup{pdfborder={1 1 1}} to extract text as a soft prompt, further refined with a hard prompt for enhanced semantic accuracy. Additionally, the SpellChecker library is employed at the receiver to detect and correct transmission-induced spelling errors caused by wireless transmission. Moreover, we employ the ControlNet-enhanced Stable Diffusion v1-5 \hypersetup{pdfborder={0 0 0}}\footnote{https://huggingface.co/lllyasviel/sd-controlnet-seg $\textnormal{\Rmnum{2}}$.}\hypersetup{pdfborder={1 1 1}}as the generative model for semantic inference at the receiver, enabling high-fidelity reconstruction and precise semantic understanding. Our experiments are conducted on a system equipped with an NVIDIA RTX 4060 GPU with 32~GB of memory, utilizing PyTorch 2.1.0 for model implementation.

\subsubsection{Network Hyperparameters}
To account for the inference time of the generative model, we limit the total number of semantic inferences to 20. We set the distances from the BS to users based on their locations and requirements, categorizing them as 30m for short-distance users, 100m for mid-distance users, and 400m for long-distance users. The other communication-related parameters and algorithm training settings are listed in Table \Rmnum{1}.
\begin{table}[!t]
\caption{Parameter settings.}
\centering
\renewcommand{\arraystretch}{0.4} % 统一行高
\setlength{\arrayrulewidth}{0.7pt} % 调整表格线条粗细
\setlength{\tabcolsep}{6pt} % 调整列间距
\begin{threeparttable}
\centering
%\begin{tabular}{|p{1cm}|p{2cm}|p{5cm}|} % 手动设置每列的宽度
\begin{tabular}{|c|c|} % 手动设置每列的宽度
\hline
\textbf{Parameters} & \textbf{Values} \\ \hline
Bandwidth & \makecell[c]{$B$ = 10~MHz}  \\ \hline
Number of users & \makecell[c]{$K$ = 3}  \\ \hline
Number of transmit antennas & \makecell[c]{$N_t$ = 8}  \\ \hline
Gaussian noise  & \makecell[c]{$\sigma_k^2=-174~\mathrm{dBm} / \mathrm{Hz}, \forall k \in \mathcal{K}$}  \\ \hline
Pathloss exponent & \makecell[c]{$\alpha$ = 3.4}  \\ \hline
Pathloss coefficient & \makecell[c]{$\eta_0$ = -30~dB}  \\ \hline
Actor Network learning rate & \makecell[c]{$\eta_0$ = 1e-3}  \\ \hline
Critic Network learning rate & \makecell[c]{$\eta_0$ = 3e-4}  \\ \hline
Discount factor & \makecell[c]{$\eta_0$ = 0.99}  \\ \hline
Clipping hyperparameter & \makecell[c]{$\epsilon$ = 0.2}  \\ \hline
Mini-batch size & \makecell[c]{128}  \\ \hline
Number of episodes & \makecell[c]{15000}  \\ \hline
\end{tabular}
\end{threeparttable}
\end{table}

\subsubsection{Comparison Scheme}
The proposed SS-MGSC framework enhances communication system performance by jointly optimizing the quantity of semantic extraction and the quality of semantic transmission, overcoming the limitation of solely optimizing SI transmission or SI encoding. To demonstrate the transmission efficiency of the proposed scheme, we delve into the role of our SS-MGSC and design the following comparison scheme:

\begin{itemize}
\item[$\bullet$] The SS-MGSC framework utilizes SSMA to simultaneously transmit both image and textual SI. To demonstrate its spectrum efficiency enhancement, we employ space division multiple access as a benchmark, transmitting image and textual SI separately for comparison. The baselines that transmit only the one-hot map and only the text are referred to as O-MGSC and T-MGSC, respectively.

\item[$\bullet$] We extract and transmit SI, significantly reducing the data volume. To demonstrate that reducing the transmitted data volume does not compromise performance, we compare it with a baseline scheme transmitting the original image without semantic extraction. We compress the original image using JPEG approximately equal to that of the transmitted one-hot maps.

\item[$\bullet$] We employ one-hot encoding to represent the SI of the image for transmission, taking advantage of its binary nature to enhance robustness against noise. To validate the noise resilience of the one-hot encoding, we also transmit semantic segmentation maps (SegS-MGSC) and evaluate the results under varying transmission power levels using the SES metric and visualized regenerated images, thereby providing further evidence of its noise-resistant properties. 
\end{itemize}

\subsection{Convergence Analysis of the SS-MGSC Framework}
\begin{figure}[!t]
    \centering
    \includegraphics[scale=0.42]{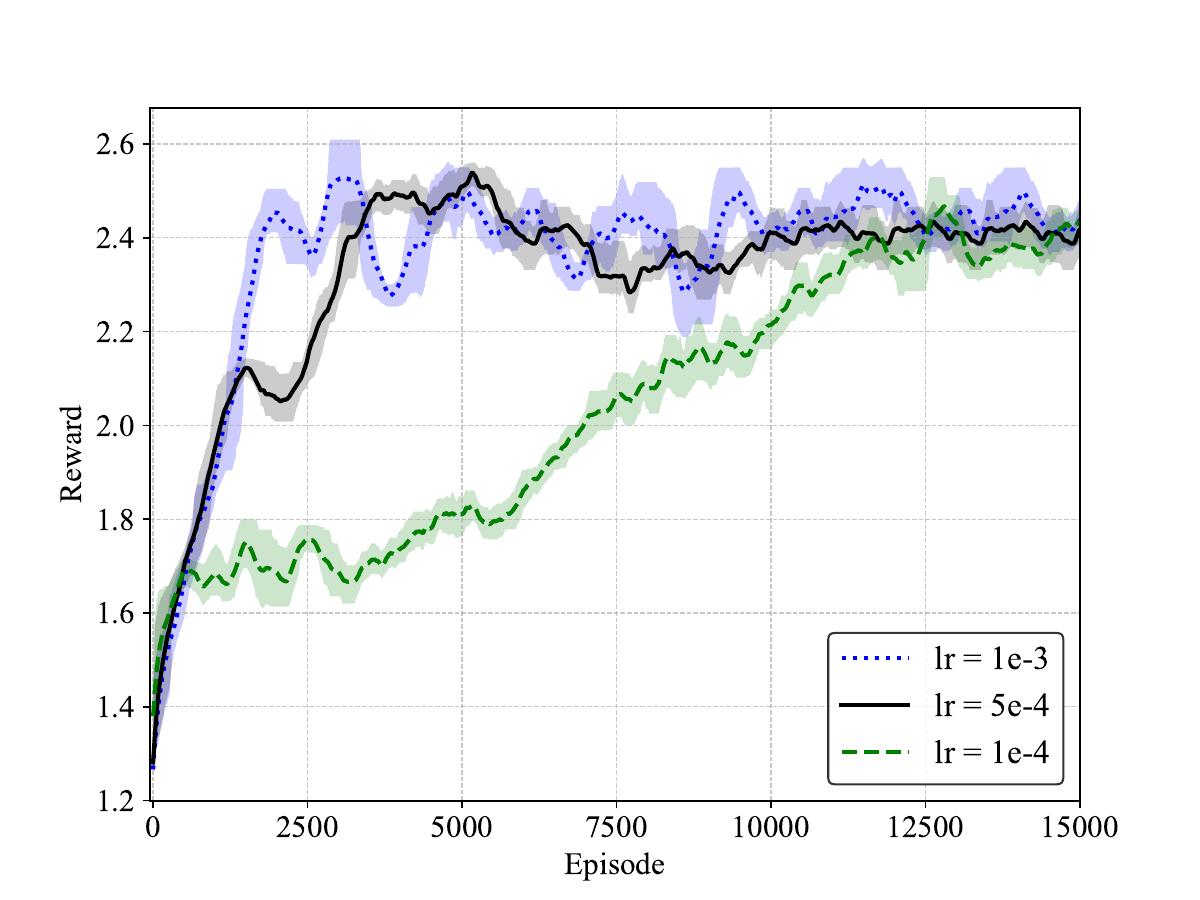}  
    \caption{Convergence performance of the PPO-based SES-oriented SSMA optimization under varying learning rates.}
    \label{Convergence lr}
\end{figure}

\begin{figure}[!t]
    \centering
    \includegraphics[scale=0.42]{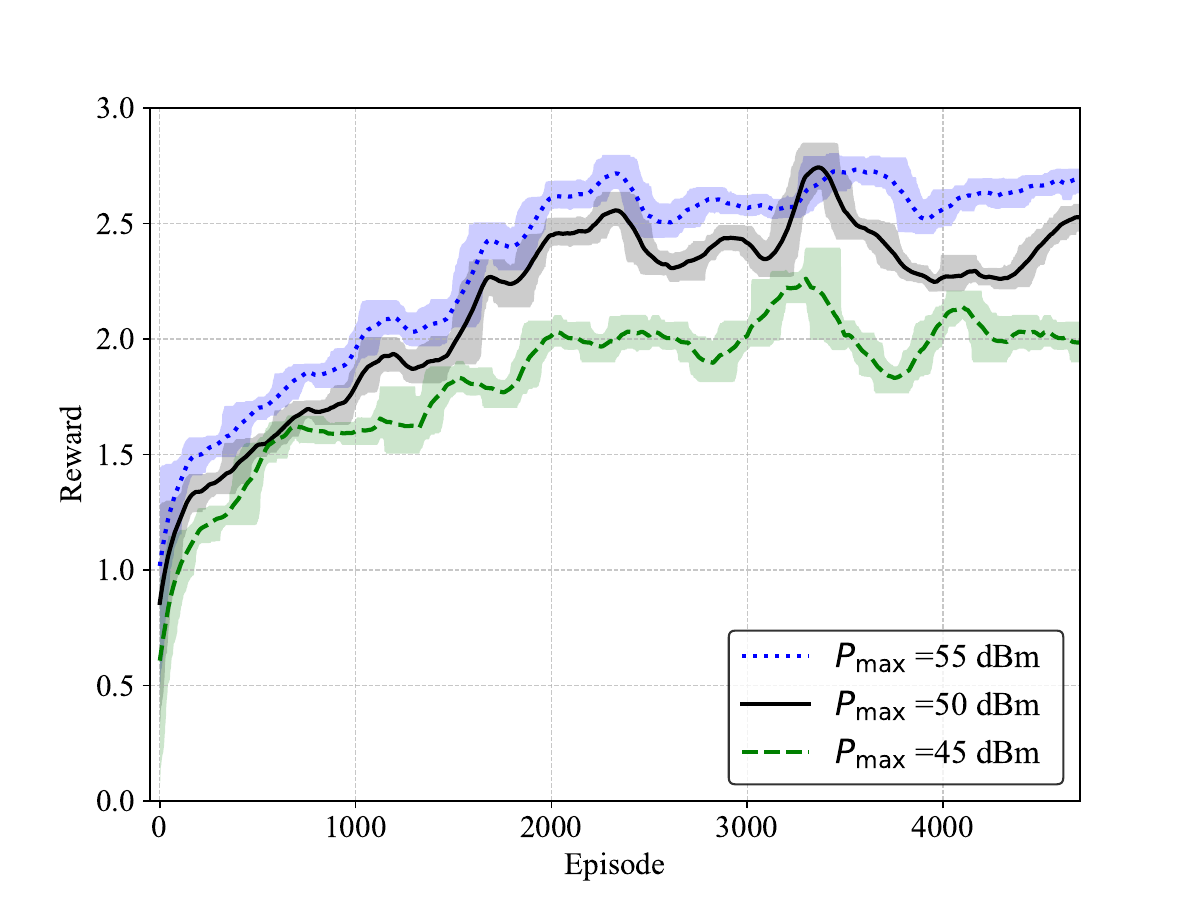}  
    \caption{Convergence performance of the PPO-based SES-oriented SSMA optimization under varying maximum transmit power.}
    \label{Convergence_P}
\end{figure}

In Fig.~\ref{Convergence lr}, we first validate the convergence performance of our proposed SS-MGSC framework. Our objective is to maximize the SES while ensuring the quality of the generated images for each user and satisfying the total transmit power constraint. The training process of
the algorithm for different learning rates is plotted. We observe that the total reward of the algorithm starts to converge at around 2,500 episodes with a learning rate of 1e-3, while a learning rate of 5e-4 achieves stable convergence at approximately 7,500 episodes. Compared to the first two learning rates, the learning rate 1e-4 results in the slowest convergence, with noticeable improvement observed only after 12,500 episodes. Despite these differences in convergence speed, all three learning rates eventually reach similar final reward values, demonstrating the effectiveness and robustness of the proposed algorithm.

Figure \ref{Convergence_P} reveals the convergence and reward performance of our proposed SS-MGSC framework with different maximum transmit power at the BS. The PPO-based SES-oriented SSMA optimization algorithm continuously interacts with the environment to learn and select optimal actions. We conduct PPO training under three maximum transmit power levels, resulting in three distinct convergence curves. We observe that the total reward converges after the same episodes across different maximum transmit power settings. Notably, the final reward value increases with higher maximum transmit power, indicating that greater transmission resources enhance 
semantic efficiency.

\subsection{Performance Analysis under Varying Maximum Transmit Power}
\begin{figure}[!t]
    \centering
    \includegraphics[scale=0.42]{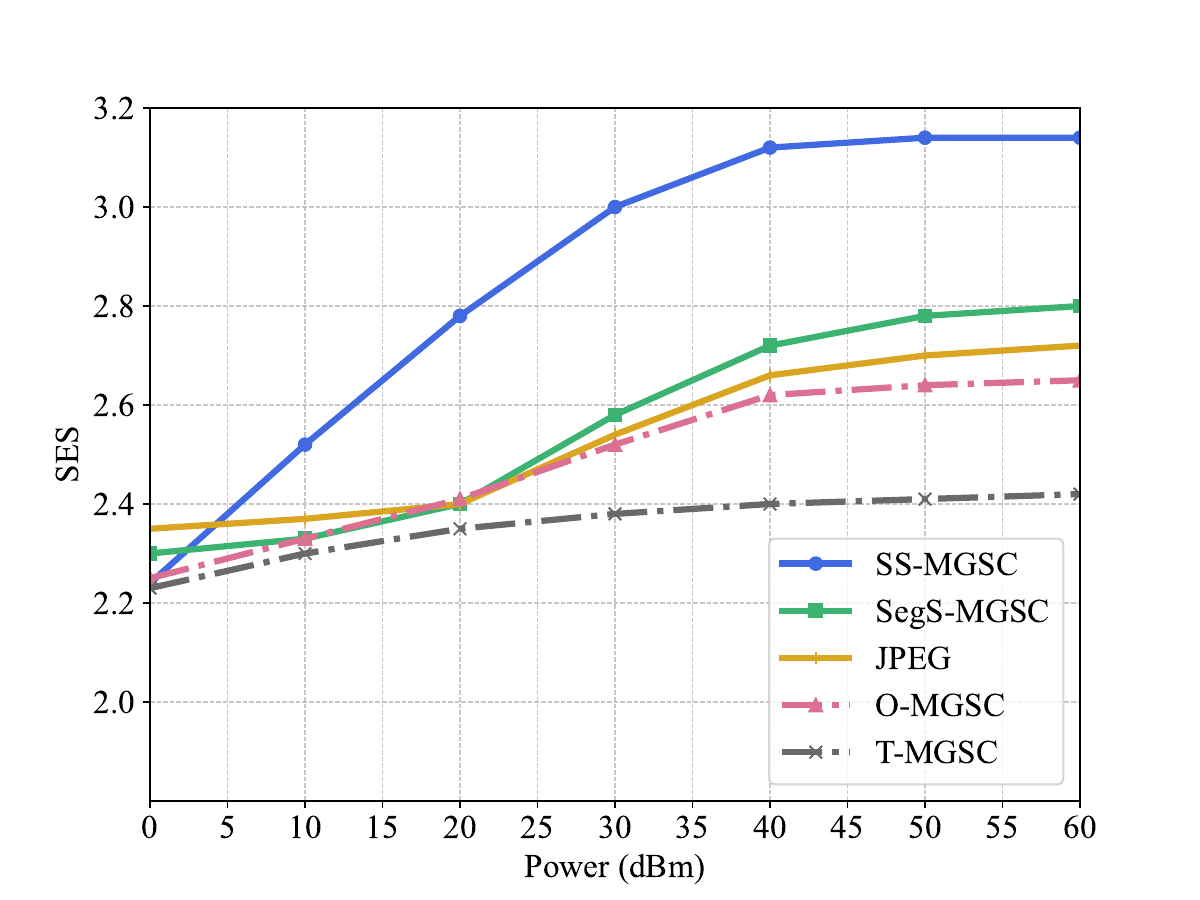}  
    \caption{Performance comparison of different schemes for PPO-based SES-oriented SSMA optimization under the same transmit power.}
    \label{SES}
\end{figure}

To demonstrate the effectiveness of the proposed SS-MGSC framework, we compare its semantic performance with that of SegS-MGSC, JPEG, O-MGSC, and T-MGSC under the same transmit power conditions. As illustrated in Fig.~\ref{SES}, SS-MGSC achieves superior semantic fidelity across different schemes. It can be clearly observed that as the transmit power increases, the SES initially exhibits a significant upward trend. However, when $P_{\max}$ exceeds 40~dBm, the improvement in SES becomes marginal. The SES performance of SegS-MGSC is lower than that of the proposed SS-MGSC, but slightly higher than that of the JPEG scheme, which demonstrates the noise robustness of the adopted one-hot map. Meanwhile, the converged total reward increases with higher transmit power levels, further confirming that greater transmission resources enhance the optimization of SI delivery. Furthermore, the lowest SES performance is observed in the O-MGSC and T-MGSC schemes, with T-MGSC yielding the poorest results among all. This finding further corroborates the effectiveness of the proposed ControlNet-enhanced Stable Diffusion, which demonstrates superior semantic efficiency performance compared to the Stable Diffusion using image-only input. These observations also indicate that visual SI may play a more pivotal role than textual input in guiding semantically faithful image generation within the reconstruction image.

\begin{figure}[!t]
    \centering
    \includegraphics[scale=0.42]{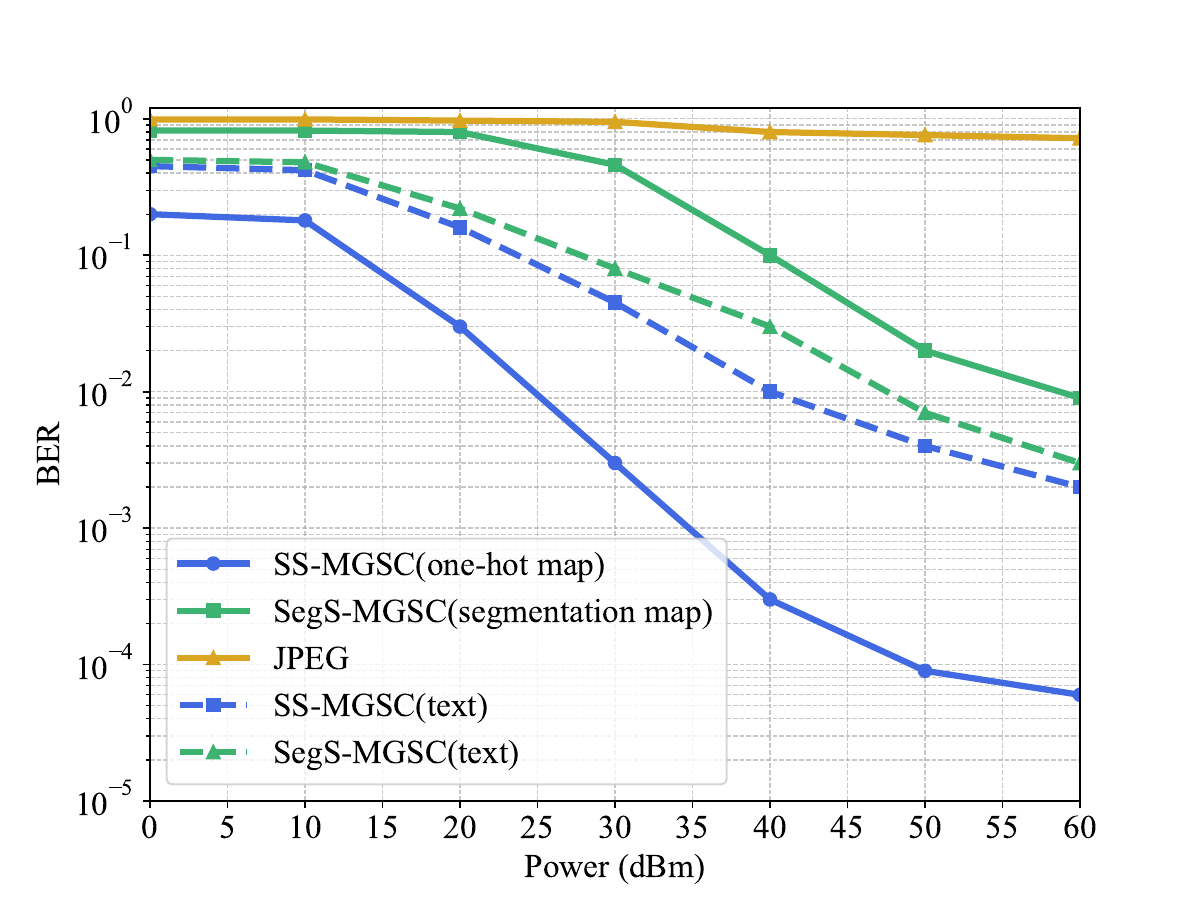}  
    \caption{BER comparison of different schemes for PPO-based SES-oriented SSMA optimization under the same transmit power.}
    \label{BER_compare}
\end{figure}

\begin{figure*}[!t]
    \centering
    \includegraphics[scale=0.23]{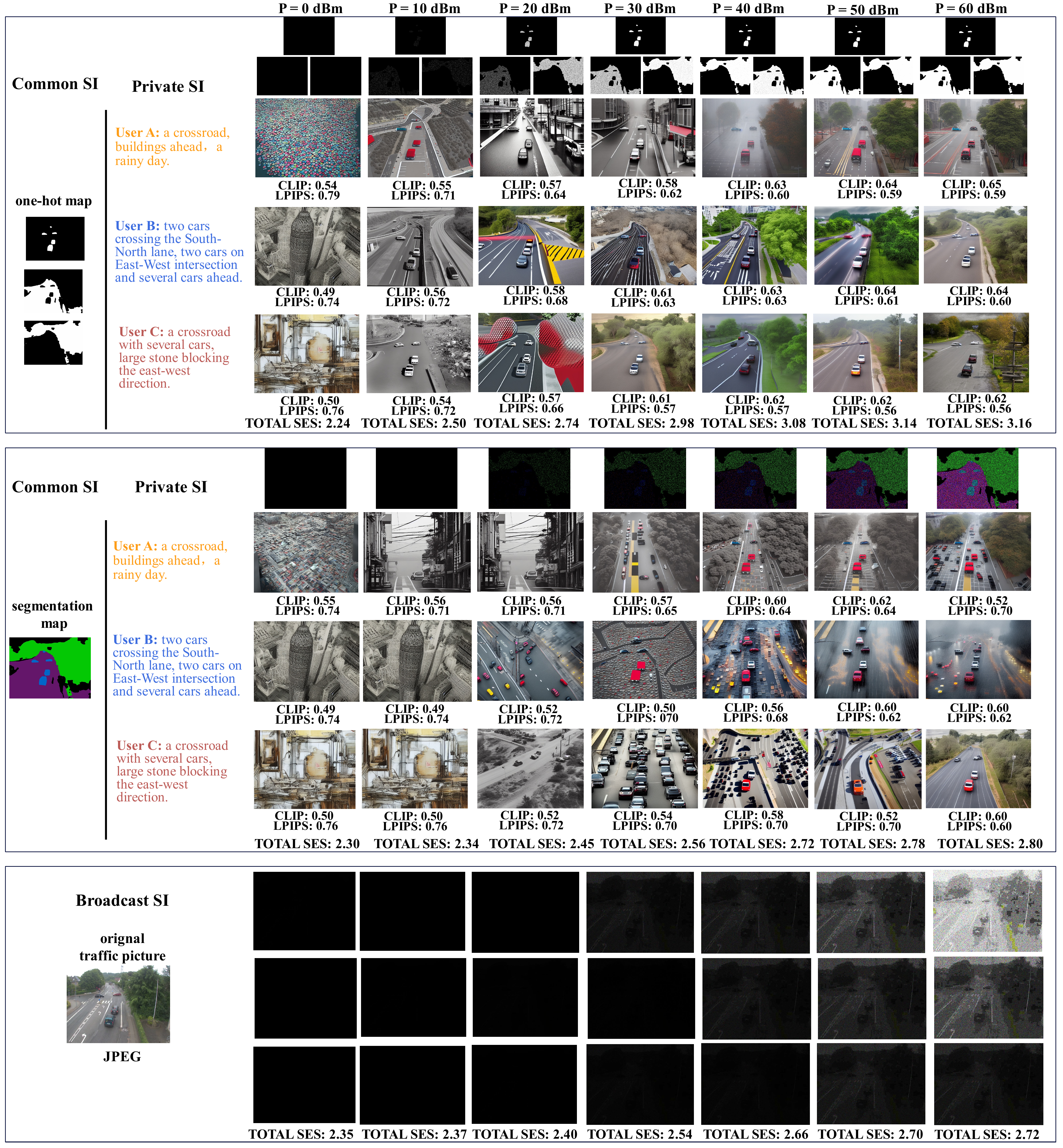}  
    \caption{Generated images under different maximum transmit power levels for SS-MGSC, SegS-MGSC, and JPEG transmission.}
    \label{generated}
\end{figure*}

To gain deeper insight into the underlying factors contributing to the SES improvement with increasing transmit power, Fig.~\ref{BER_compare} illustrates the BER performance of SS-MGSC, SegS-MGSC, and JPEG under different maximum transmit power. For SS-MGSC and SegS-MGSC, where image and textual SI are transmitted separately, the BERs of the image and text streams are reported independently to facilitate a more granular performance comparison. As shown in the figure, all schemes exhibit relatively high BERs when the maximum transmit power $P_{\max}$ is below 10~dBm. With the increase of transmit power, the BER gradually decreases. However, when $P_{\max}$ exceeds 40~dBm, the rate of decline becomes less pronounced, indicating a saturation effect in error performance. SS-MGSC consistently have the lowest BER, with the image stream exhibiting fewer errors than the text stream. This aligns with the results in Fig.~\ref{SES}, where SS-MGSC attains the highest SES across all transmit power. In comparison, SegS-MGSC yields higher BER than SS-MGSC but lower than JPEG. Unlike SS-MGSC, its text BER is lower than that of the image, which may be attributed to the lower noise robustness of segmentation maps compared to one-hot map, as well as a greater disparity in data size between image and text.

Fig.~\ref{generated} illustrates the visual results of generated images for users in three different scenarios under varying transmit power across the SS-MGSC, SegS-MGSC, and JPEG schemes. As transmit power increases, the clarity of the transmitted one-hot map and segmentation maps improves, resulting in progressively enhanced structural consistency and semantic fidelity in the generated images. It is evident that, with increasing transmit power, the SS-MGSC scheme exhibits a rise in CLIP similarity scores and a corresponding decline in LPIPS values. Under the SS-MGSC scheme, the visual differences between images generated at $P_{\max}$ = 40~dBm, $P_{\max}$ = 50~dBm, and $P_{\max}$ =  60~dBm are negligible, indicating that lower transmit power can be employed to conserve energy without sacrificing generation quality. Compared to SegS-MGSC, where the generated image exhibits structural distortion with inaccurate vehicle positions and counts, SS-MGSC yields notably superior image quality. It also demonstrates substantial improvements over the JPEG-based method in terms of both perceptual realism and semantic similarity. Under the same base station transmit power, SS-MGSC is able to deliver higher-quality images with fewer resource requirements, highlighting its efficiency in semantic transmission. Furthermore, in the proposed SS-MGSC scheme, noticeable differences in the generated images of Users A, B, and C can be observed at $P_{\max}$ = 40~dBm. User A’s image predominantly depicts road and weather conditions, User B’s highlights vehicle status, and User C’s focuses on the surrounding road environment. These differences are attributed to the private SI transmitted by the BS, further validating that the text prompts effectively fulfill user-specific detail requirements.

\subsection{Performance Analysis of SES under Varying BER Levels}
\begin{figure}[!t]
    \centering
    \includegraphics[scale=0.42]{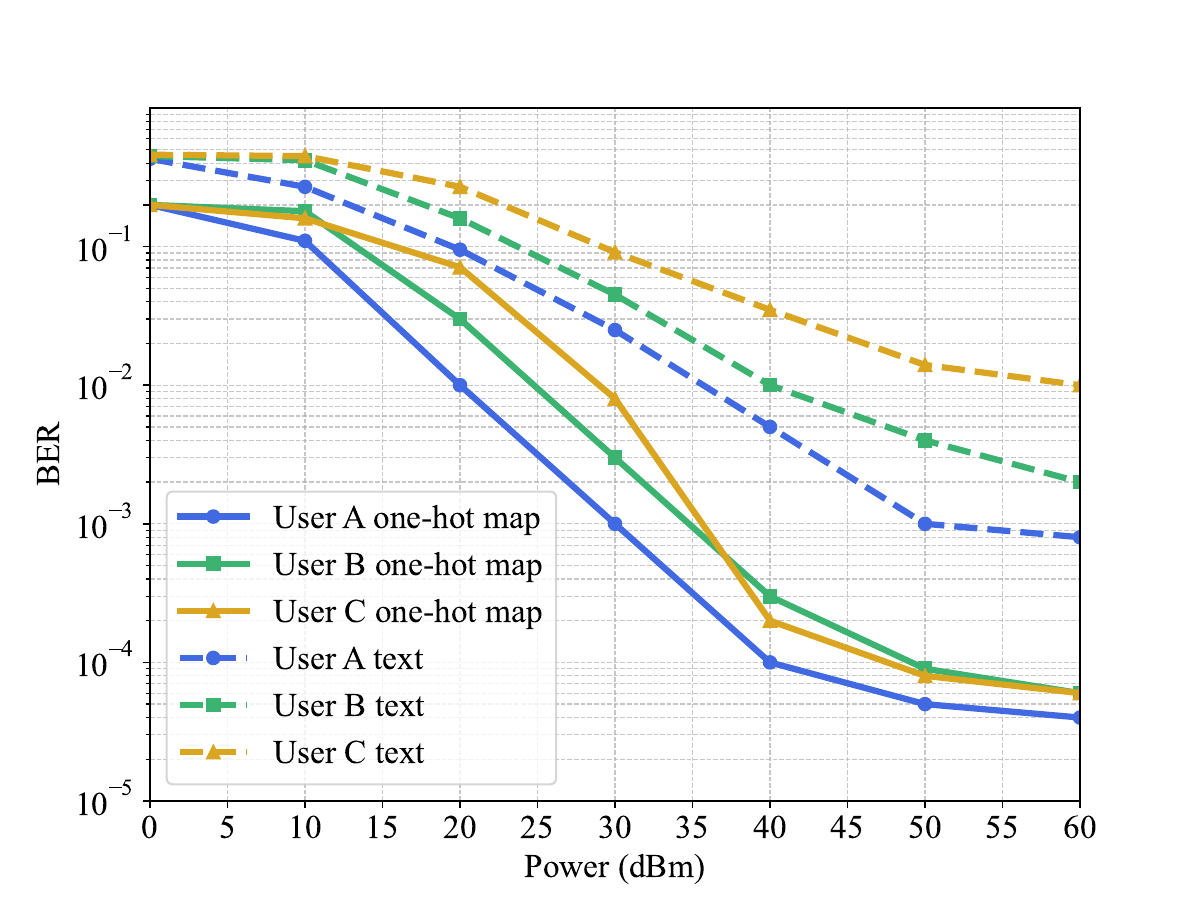}  
    \caption{BER comparison of image and text transmission in the SS-MGSC framework for different users given the same transmit power.}
    \label{BER_user}
\end{figure}

\begin{figure}[!t]
    \centering
    \includegraphics[scale=0.42]{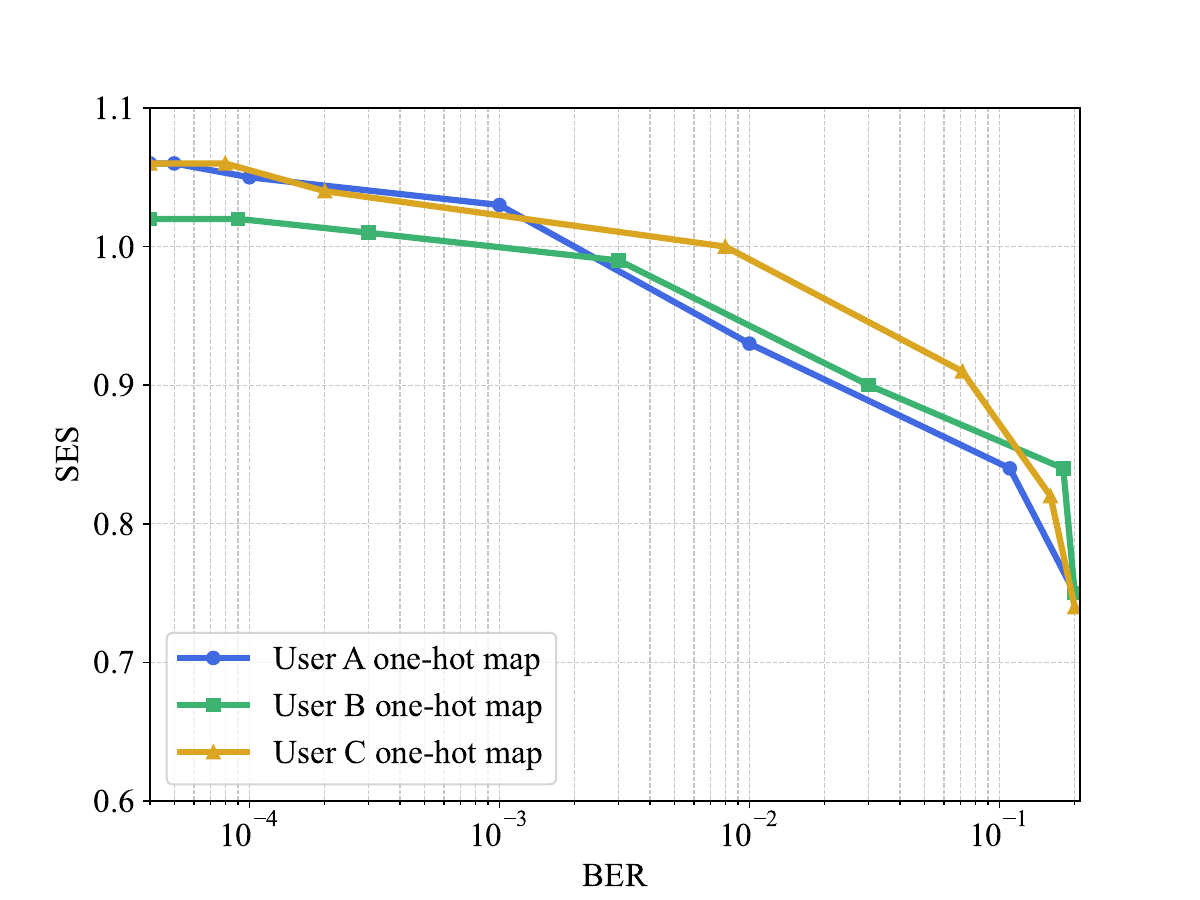}  
    \caption{SES comparison for different users under image BER in the SS-MGSC framework.}
    \label{BER_SES_onehot}
\end{figure}

\begin{figure}[!t]
    \centering
    \includegraphics[scale=0.42]{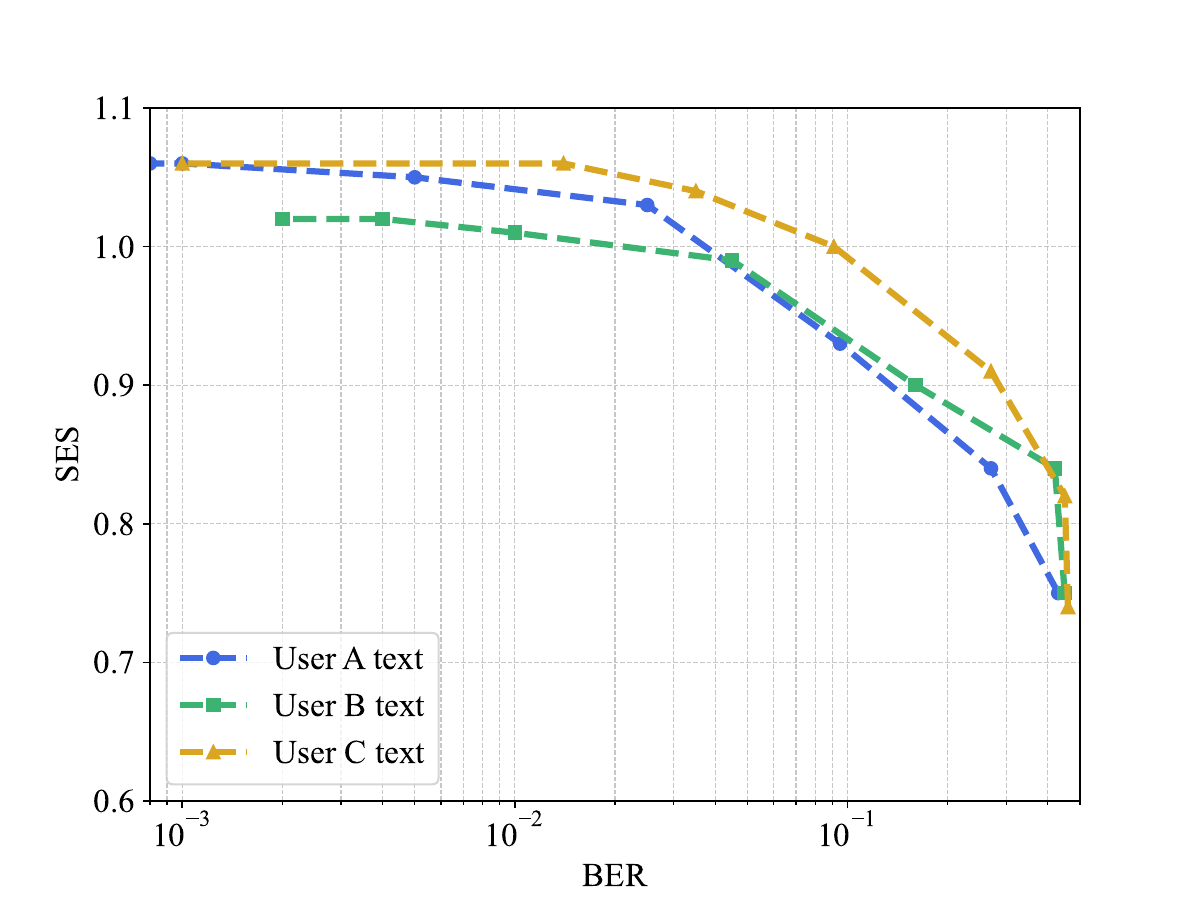}  
    \caption{SES comparison for different users under text BER in the SS-MGSC framework.}
    \label{BER_SES_text}
\end{figure}

We further investigate the relationship between BER and SES in the SS-MGSC framework. Specifically, for text transmission and one-hot map transmission, Fig.~\ref{BER_user} presents the BER performance of different users under varying transmit power. It can be observed that the BER of text transmission is generally higher than that of one-hot transmission, further validating the superior noise robustness of the one-hot encoding scheme. In addition, a clear downward trend in BER is observed between $P_{\max}$=10~dBm and $P_{\max}$=40~dBm, indicating the effectiveness of increased transmit power in reducing transmission errors within this range.

Furthermore, building upon the previously observed relationship between SES and transmit power, we further present the relationship between SES and image BER in Fig.~\ref{BER_SES_onehot} and that between SES and text BER in Fig.~\ref{BER_SES_text}. As shown in the figures, users in three scenarios exhibit similar trends. Specifically, when the BER exceeds 1e-3,  the SES experiences a significant decline. This observation is consistent with the earlier trend observed at $P_{\max}$=40~dBm, where the higher BER levels corresponded to a noticeable drop in SES performance.

\section{Conclusion}
In this work, we have addressed the redundancy problem in multi-user content distribution by proposing a novel framework for multi-user communication systems, referred to as SS-MGSC. To validate the effectiveness of the proposed framework, we have considered V2X communication as a representative use case. Within SS-MGSC, we have jointly optimized beamforming and the common and private SI, aiming to maximize semantic efficiency while minimizing resource consumption. In addition, we have constructed a SKB using intent-aware semantic parsing, which integrates both environmental conditions and user preferences. Furthermore, we have designed a novel semantic evaluation metric, SES, which not only ensures semantic similarity but also preserves perceptual structural consistency in the generated images. Next, we have proposed RL-based algorithms to jointly optimize power allocation and the selection of common and private SI. Simulation results have demonstrated the effectiveness of the proposed architecture, which plays a pivotal role in supporting road condition–related tasks in vehicular networks. Among all evaluated schemes, SS-MGSC has consistently achieved the superior performance, highlighting its superiority in SemCom. Future work will focus on further exploring GAI-enhanced SemCom, with particular attention to advancements at the transmitter side and within the channel transmission process. In addition, we will work on reducing the computational overhead at the receiver introduced by ControlNet-enhanced diffusion models and mitigating the system’s sensitivity to channel estimation errors.

\vfill

\end{document}